# Comment on "Thermodynamic determination of the equilibrium first-order phase-transition line hidden by hysteresis" Sci Rep **13**, 6876 (2023); arXiv:2303.00327, by K. Matsuura et al.

P Chaddah*

UGC-DAE Consortium for Scientific Research

Indore 452001 India

*We stress that the lower hysteresis line should not be used to infer any thermodynamic quantity below the temperature at which the lower hysteresis line shows a maximum.*

The following two results of Matsuura et al [1] are being commented on,

1. The authors stress that for a field-induced FOT exhibiting large hysteresis, the line of the mid-points of the lower-field and higher-field boundaries of the hysteresis region, is not the equilibrium phase-transition line.

2. If this line of mid-points viz. $H_0(T)$ is used as the equilibrium phase-transition line in their study on $Fe(Zn)_2Mo_3O_8$ then something peculiar is found from the Clausius-Clapeyron equation below 18K. Specifically, the entropy difference (and the magnetization difference) between the two phases would show complicated behaviour, and would switch its sign below this temperature. Recognizing that such an unusual behaviour requires novel origin like in the melting curve of helium-3, they consider this an unphysical artefact arising of an erroneously inferred equilibrium phase-transition line. The equilibrium phase transition line determined through their studies, that avoid the "two-phase mixture" region of field and temperature, flattens below 18K and does not show peculiarities below this temperature. The lower-field boundary of hysteresis, they conclude, should not be used below 18K.

On the first point we note that the hysteresis in temperature-dependent measurements is dictated by the limits of supercooling (T*) and superheating (T**). At these thermodynamic stability limits the metastable states become unstable [2]. The actual observed hysteresis could be smaller because fluctuations cause the transition even while a finite barrier of free-energy separates the metastable state from the stable state.

We take two examples of free-energy functions that are in common use [3,4]. Following sections 2.3 and 2.4 of reference [4], we note that for a form of the free energy that is symmetric under inversion of the order parameter, we have $T_C - T^* = 3[T^{**}-T_C]$ ; and for a form of the free energy that is not symmetric under inversion of the order parameter, we have $T_C - T^* = 8[T^{**}-T_C]$ . In both cases we note that $T_C \neq [T^* + T^{**}]/2$ , with $T_C$ being closer to T** than to T*.

We thus note that $T_C$ will not, except in some fortuitous case, be the mean of T* and T**. One therefore cannot assert that $T_C$ is the mid-point of the lower-temperature and higher-temperature boundaries of the hysteresis region. By analogy, one also cannot assert that $H_0$ is the mid-point of the lower-field and higher-field boundaries of the hysteresis region. The mid-point of hysteresis being taken as the equilibrium phase-transition point is an approximation frequently used by experimentalists, but is not rigorously valid even in simple models for first-order transitions.

We now come to the second point of peculiar behaviour of the hysteresis mid-point at low temperature. We wish to assert that the peculiarity is actually in the behaviour of the lower-field limit of the hysteresis, which starts reducing with lowering temperature below 30K, while the upper-field limit does not show such peculiarity and is continuously increasing with reducing temperature. We assert that the lower-field limit of the hysteresis should not be used below 30K (i.e. even above 18K) as it corresponds to the de-arrest of the higher magnetization FRI phase that has undergone glass-like kinetic arrest in this material. It is not due to the thermodynamic phase transformation from metastable FRI phase to stable AFM phase.

Matsuura et al [1] note that the hysteresis mid-point $H_0$ reduces with lowering temperature below 18K, and using $dH_0/dT$ with Clausius-Clapeyron equation implies that the transformation entropy change becomes negative. Matsuura et al term this as a complicated behaviour which, they show through their detailed analysis for extracting the equilibrium phase transition line, is not correct.

Umetsu et al [5,6] reported similar peculiar behaviour in Ni-Mn-In in the hysteresis mid-point in magnetization [5], and in the hysteresis mid-point in resistivity [6], at temperatures below about 50K under isothermal variation of magnetic field. Umetsu et al [5,6] conclude in both papers that the transformation entropy change becomes negative at low temperature. They were aware of the phenomenon of kinetic arrest but stressed that they consider it to be thermal transformation arrest, and probably did not consider it while they were studying magnetic-field-induced transformations.

Such peculiarity in the lower-field limit of the hysteresis was observed and highlighted as a consequence of glass-like ***kinetic arrest*** of the first-order magnetic transition in both the half-doped manganite $Nd_{0.5}Sr_{0.5}MnO_3$ [7] and in FeRh(Pd) that also show a magnetic first order transition [8]. As discussed in sections 4.3 and 5.3 of ref [4] while discussing manifestations of kinetic arrest during isothermal variation of magnetic field, the 'lower hysteresis line' is observed as the higher field kinetically arrested phase is de-arrested with lowering field. The high-field arrested phase transforms to the lower-field de-arrested equilibrium phase, and magnetization and resistivity show a jump under isothermal reduction of magnetic field. But this part of the 'lower hysteresis line' is dictated by kinetics, and does not correspond to thermodynamic processes. Figure 5 of reference [7] and figure 6 of reference [8] show results of isothermal field cycling in these two different magnetic materials, along with the insets in each figure that explain the lower temperature non-monotonicity (similar to reports in ref [1] and ref [5,6]) as due to the lower hysteresis line being the kinetic arrest line.

Thermodynamics looks at transitions that occur at equilibrium, when the free energy of the two phases are equal. As noted in this paper, thermodynamics allows transitions to occur away from equilibrium and up to the metastability limit or the spinodal, from a metastable state to the equilibrium state and in such cases hysteresis is observed. Transition between two phases, with sharp change in physical properties, can also be observed beyond the metastability limit when a glass-like arrested state is de-arrested. Hysteresis in the isothermal scan will again be observed. We note that the lower field of the hysteretic scan shifts from the T* spinodal to the de-arrest line with lowering temperature, and will go from increasing with reducing T to decreasing for reducing T. We assert that the data of Matsuura et al [1], and also the earlier data of Umetsu et al [5,6], is in conformity with this explanation.

It follows that the lower hysteresis line, below the temperature $T_{max}$ at which its maximum is observed, is a manifestation of the arrested high magnetization FM phase being de-arrested and transforming to the equilibrium AFM phase. It has no relation to the equilibrium phase transition line. The upper hysteresis line could be used below $T_{max}$ by doing an analytic continuation of its relation to the equilibrium phase transition line obtained at higher temperature. The kinetic arrest, and de-arrest, lines are dictated by kinetics. These do not correspond to thermodynamic processes.

While Matsuura et al do not use the mid-point [$H_0$] below the temperature (18K) at which the mid-point shows a maximum, we stress that the lower hysteresis line should not be used to infer any thermodynamic quantity below the temperature (30K) at which this lower hysteresis line shows a maximum.

*\*Since retired*